# Symmetry selective third harmonic generation from plasmonic metacrystals

Shumei Chen[1,2*], Guixin Li[1,2*], Franziska Zeuner[3*], Wing Han Wong[4], Edwin Yue Bun Pun[4], Thomas Zentgraf[3†], Kok Wai Cheah[2‡], Shuang Zhang[1§]

[1]School of Physics & Astronomy, University of Birmingham, Birmingham, B15 2TT, UK
[2]Department of Physics, University of Paderborn, Warburger Straße 100, D-33098 Paderborn, Germany
[3]Department of Physics, Hong Kong Baptist University, Kowloon Tong, Hong Kong
[4]Department of Electronic Engineering, City University of Hong Kong, 83 Tat Chee Ave, Hong Kong

**Abstract: Nonlinear processes are often governed by selection rules imposed by the symmetries of the molecular configurations. The most well-known examples include the role of mirror symmetry breaking for the generation of even harmonics, and the selection rule related to the rotation symmetry in harmonic generation for fundamental beams with circular polarizations. While the role of mirror symmetry breaking in second harmonic generation has been extensively studied in plasmonic systems, the investigation on selection rules pertaining to circular polarization states of harmonic generation has been limited to crystals, i.e. symmetries at the atomic level. Here we demonstrate the rotational symmetry dependent third harmonic generation from nonlinear plasmonic metacrystals. We show that the selection rule can be imposed by the rotational symmetry of meta-crystals embedded into an isotropic organic nonlinear thin film. The results presented here may**

---

[*] These authors contributed equally to this work.

[†] thomas.zentgraf@uni-paderborn.de

[‡] kwcheah@hkbu.edu.hk

[§] s.zhang@bham.ac.uk

open new avenues for designing symmetry-dependent nonlinear optical responses with tailored plasmonic nanostructures.

In conventional nonlinear optics, the selection rules of harmonic generation and other nonlinear optical processes are mainly based on the symmetries at the most fundamental levels, such as of molecular structures and atomic crystal configurations[1, 2]. However, the macroscopic symmetry of photonic crystals, nanoparticles, quantum dots, and liquid crystals may lead to new selection rules for the nonlinear optical processes. The importance of macroscopic symmetry is especially prominent in plasmonic nanostructures, in which localized surface plasmon polariton excitations can be very sensitive to shape, size, and surrounding medium[3]. The strong field localization in plasmonic nanostructures has been utilized to enhance the second harmonic generation (SHG) [4-14], third harmonic generation (THG) [15-18], and four-wave mixing[19-21]. In the past several years, much work has been dedicated to the investigation of SHG in plasmonic structures with broken mirror symmetry, such as split ring resonators[4], L and V-shaped nanorods[5, 6], and G-type chiral metamolecules[7, 8]. Very recently, symmetry related selection rules were investigated in chiral metamaterials with four-fold rotational symmetry. It was found that the four wave mixing only occurs when the signal beam and the pump beam have the same circular polarization, and the resulting nonlinear generated beam has the same circular polarization as well[22].

It was discovered that harmonic generation with circularly polarized light obeys certain selection rules which relate the order of harmonic generation to the rotational symmetry of the molecules[23-25]. The selection rules state that for a circularly polarized fundamental wave propagating along the axis of a crystal or molecule with $n$ fold rotational symmetry, the allowed orders for harmonic generation are $l(n\pm1)$, with $l$ being an integer, and the '+' and '-' sign corresponding to harmonic generation of the same or opposite circular polarization state, respectively. Based on the selection rules for THG, circularly polarized (CP) light cannot generate a third harmonic wave in isotropic media[23-27], since the spin angular momentum of the light would not be conserved in such processes. One interesting property of THG with circular polarization is that CP light produces THG with only opposite handedness when it propagates along the axis of a crystal with four-fold rotational symmetry[26, 27]. This rule holds for the microscopic molecular symmetry, however, it has not been experimentally verified at the mesoscopic symmetry level of artificial nanostructures, where the significant enhancement of the field due to the strong interaction between light and the nanostructure greatly increases the efficiency of the nonlinear processes.

For the investigation of the correlation between rotational symmetry and third harmonic generation of circularly polarized light, we designed three types of plasmonic crystals consisting of an array of rectangular nanorods with C2 rotational symmetry, cross-nanorods with C4 symmetry, and three pedal nanorods with C3 symmetry (Fig.1a). Note that the nanostructures with C2 and C4 symmetries are arranged in a square lattice, whereas structures with C3 symmetry are arranged in a triangular lattice, such that the symmetries of the structures and the lattice are consistent with each other. A thin organic film made of poly(9,9-dioctylfluorence) (PFO) which has a high third-order nonlinearity was coated onto the plasmonic crystals to generate the THG in the experiments. We confirmed that a bare PFO film does not produce THG under CP light excitation as predicted by the selection rules. However, CP light induced THG can be observed by introducing the C2 and C4 symmetry plasmonic crystals while it cannot be observed from samples with C3 symmetry. Hence our experiments directly prove that the selections rules for CP light induced THG also works for plasmonic nanostructures with mesoscopic symmetry. We further observe that while the rectangular nanorods with C2 symmetry generate THG light with both circular polarizations, the nanostructures with C4 symmetry produce THG with only the opposite circular polarization state (Fig. 1b). These findings provide us with more freedom to manipulate optical spin-dependent nonlinear optical processes using plasmonic macromolecules.

For our studies on the THG selection rules in plasmonic nanostructures we fabricated gold nanostructures with C2, C3 and C4 symmetries by using electron beam lithography and metal lift-off process. The geometries of the samples are chosen such that all samples show similar plasmonic resonance wavelengths in the near-infrared spectral range. Subsequently, a 100-nm-thick PFO film from a 15 mg/mL toluene solution was spin-coated on top of the gold nanostructures to form a gold-PFO hybrid plasmonic metacrystals. The scanning electron micrographs of the samples with the three different symmetries are shown in Fig. 2 (right).

For normal light incidence the linear optical properties of the samples are characterized by measuring the polarization dependent transmission spectra using Fourier transform infrared spectrometry (Fig. 2). The transmission dips correspond to the excitation of localized surface plasmon polariton (LSPP) resonances supported by the finite geometry of the nanostructures. For the C2 symmetry sample, the LSPP resonance is around 1269 nm in the near infrared for illumination with horizontal (H) polarization (along the long axis of the rod), and in the visible for vertical (V) polarized illumination (perpendicular to the rod). For the C3 symmetry sample, the LSPP resonances are at 1250 nm and 1260 nm respectively for H and V polarizations. For the C4 symmetry device, ideally, the LSPP resonances for H and V polarization should occur at the same wavelength. However, the measured resonance wavelengths are slightly different for the two linear polarizations (1207 nm for H polarization and 1219 nm for V polarization) due to the imperfection of fabrication processes. For all the samples, the numerically simulated spectra closely match the measurements.

The third harmonic generation measurements (Fig. 3a) were performed using the light from a femtosecond (fs) laser pumped optical parametric oscillator (repetition frequency: 80 MHz, pulse duration: ~200 fs, averaged power: P = 8 mW) at a wavelength of $\lambda$ =1.25 μm, which is close to the LSPP wavelength of all three plasmonic meta-crystals. In the measurement, the fundamental pulses are focused to a spot with diameter of ~ 50 μm using an infinity-corrected objective lens (5x, N.A = 0.10) with focal length of 25.8 mm. The resulting THG signal from the gold-PFO hybrid samples is collected in transmission by a second infinity-corrected objective lens (5x, N.A = 0.10) and imaged onto a color charge coupled device (CCD) camera after filtering out the fundamental wavelength using a band-pass filter (315 nm-710 nm).

The CCD images of the THG signal at a wavelength of 417 nm from the C2, C3 and C4 symmetry samples are shown in Fig. 3b, and the measured intensities of the THG signal are summarized in Table 1. As expected, we observe a THG signal from all three samples when illuminated with linear polarized light as no selection rule prohibits the THG for linear polarized light. Compared to the THG from a pure PFO reference film, the LSPP resonance of the gold nanorods dramatically improves the efficiency of THG from gold-PFO hybrid structures due to the near field enhancement of the electric field in the vicinity of the nanorods. For the C2 symmetry metacrystal, the THG intensity for H polarization is much higher than that for V polarization. This is reflected from the linear optical property of the C2 nanorod which shows that the excitation of the vertical polarization state is non-resonant (see Fig. 2b). For samples with C3 and C4 symmetries, the THG for both H and V polarizations are of similar intensities as both of the two polarization states can excite the LSPP resonance at the fundamental frequency in the near-infrared. We also observe the THG signal from a surface with only gold nanorods with no embedded PFO film is much weaker than that from a bare PFO film. This indicates that the THG of the gold-PFO hybrid systems is dominated by the contribution from the PFO film due to its large third harmonic nonlinear coefficient.

In the next step we carried out the THG measurements for circularly polarized light (Fig. 3b). For the bare PFO film we do not observe any THG signal within our detection limit when the sample is illuminated with circularly polarized light. This result agrees with the selection rules for THG; in fact the thin PFO film on the glass substrate with the organic molecules randomly oriented can be treated as an isotropic medium[22-26]. However, once the plasmonic nanorods are embedded into the PFO film, they significantly change the local field distribution and thus alter the THG process. Besides the strong field enhancement at the fundamental wavelength, the polarization state of the local electric field in the vicinity of the plasmonic structures is strongly modified by the presence of the LSPP resonance mode and is in general not circularly polarized. This gives rise to nonzero local third-order nonlinear polarization, whose emission into the far field is further mediated by the response of the plasmonic structures at the THG wavelength. Interestingly, although the local third-order nonlinear polarization is nonzero, the overall THG from the C3 symmetry sample vanishes in the far field. This is dictated by the rotational symmetry of the nanostructures and the lattices.

More intriguing results come from the comparison of the THG signals between the C2 and C4 samples. Strong optical spin dependent THG signals are observed when the circularly polarized fundamental wave is incident onto the C2 and C4 symmetry samples. For the C2 symmetry sample, right circularly polarized light illumination generates THG signals with both right circular polarization (RCP) and left circular polarization (LCP), as both are not forbidden by the selection rules. However, the RCP component of the THG signal is much stronger than the LCP component, with a polarization ratio between the LCP and RCP intensities of $\eta_{C2}$ (THG$_{LCP}$/THG$_{RCP}$) = 0.121. Interestingly, for the C4 symmetry sample, the LCP component dominates in the RCP induced THG signals, with an extremely large polarization ratio $\eta_{C4}$ (THG$_{LCP}$/THG$_{RCP}$) of ~ 29. According to the selection rule for C4 rotational symmetry, the THG light should be solely in the LCP state and THG of the right circularly polarized light should be forbidden; however, the imperfectness of the nanostructures arising from the fabrication slightly breaks the rotational symmetry and leads therefore to a small THG signal in the RCP state. Hence, all the experimental observations are consistent with the selection rules of THG with circularly polarized light excitation in crystal optics[27].

The selection rules for THG can be understood through the symmetry consideration of the field distribution of both the fundamental wave and the THG wave in the nonlinear plasmonic system. Due to the deep subwavelength thickness of the plasmonic structures, the phase matching condition along the propagation direction is not important, therefore the conversion to the THG signal can be simply expressed as[22]

$$A_\beta^{3\omega} = i\Gamma_{\beta\alpha\alpha\alpha}(A_\alpha^\omega)^3 \qquad (1)$$

whereas the indices $\alpha, \beta \in \{\text{RCP}, \text{LCP}\}$ denote the right or left circular polarizations, $i$ the imaginary unit, and $A$ the amplitude of the fundamental and THG waves, respectively. $\Gamma$ is the nonlinear conversion coefficient and is given by

$$\begin{aligned}\Gamma_{RRRR} &\propto \int dV \left[ \chi_{loc}^{(3)}(\vec{r})\vec{e}_R^\omega(\vec{r})(\vec{e}_R^\omega(\vec{r})\cdot\vec{e}_R^\omega(\vec{r})) \right] \cdot \vec{e}_R^{3\omega}(\vec{r})^* \\ \Gamma_{LRRR} &\propto \int dV \left[ \chi_{loc}^{(3)}(\vec{r})\vec{e}_R^\omega(\vec{r})(\vec{e}_R^\omega(\vec{r})\cdot\vec{e}_R^\omega(\vec{r})) \right] \cdot \vec{e}_L^{3\omega}(\vec{r})^* \end{aligned} \qquad (2)$$

Here $\chi^{(3)}$ is the local third-order susceptibility of PFO and gold. $\vec{e}_R^\omega(\vec{r})$ is the normalized electric field distribution of the fundamental wave excited by a RCP incident beam of unit amplitude, and $\vec{e}_{R(L)}^{3\omega}(\vec{r})$ is the field distribution of the THG in the plasmonic structure responsible for generating a right (left) handed circularly polarized THG wave in the transmission direction. Note that, due to the time reversal connection between a RCP wave propagating in the forward direction and a LCP wave propagating in the backward direction, $\vec{e}_{R(L)}^{3\omega}(\vec{r})$ can also be considered as the field distribution in the plasmonic structure when it is excited by an incident LCP (RCP) wave at the THG frequency and unit amplitude along the opposite direction to that of the incident fundamental wave.

For a plasmonic structure of *n*-fold rotational symmetry it is straightforward to show from Eq. (2) that applying the rotation operation of 2π/*n* on the nonlinear conversion coefficient leads to:

$$R_{2\pi/n}(\Gamma_{RRRR}) = \exp(i\frac{4\pi}{n})\Gamma_{RRRR}$$

$$R_{2\pi/n}(\Gamma_{LRRR}) = \exp(i\frac{8\pi}{n})\Gamma_{LRRR}$$

(3)

The above equations result from the rotational operation on the field distribution excited by the circular polarization $R(\vec{e}_R(\vec{r})) = \exp(i\frac{2\pi}{n})\vec{e}_R(\vec{r})$ and $R(\vec{e}_L(\vec{r})) = \exp(-i\frac{2\pi}{n})\vec{e}_R(\vec{r})$, respectively. Due to the rotational symmetry of the plasmonic nanostructures, the rotation operation by $2\pi/n$ should keep the nonlinear conversion coefficient unchanged. Thus, THG can only be generated in the far-field if the exponential term in Eq. (3) equals to unity. For a right circularly polarized incident beam, THG with RCP can be generated only for $n = 2$, whereas THG with LCP can exist for $n = 2$ and $n = 4$. The simple symmetry based analysis nicely explains our experimental observation of THG on plasmonic structures with C2, C3 and C4 rotational symmetries: THG is forbidden in the C3 structure, and allowed in C2 and C4 structures. Most interestingly, for the C4 structure, only THG of the opposite circular polarization to that of the incident beam is observed.

In conclusion, we have investigated the impact of rotational symmetry on the third harmonic generation for circularly polarized light in nonlinear plasmonic crystals. We showed that THG with a circularly polarized incident beam can only be generated in plasmonic structures with two- and four-fold rotational symmetries, but is forbidden in the three-fold rotational symmetry configuration. The experimental observations on the three types of plasmonic structures exhibit selection rules that have been observed previously only at the atomic crystal or molecular levels. Our study paves the way towards the design of novel spin dependent nonlinear plasmonic devices.

**Acklowledgment**

This work was partly supported by EPSRC. T.Z. and S.Z. acknowledge financial support by the European Commission under the Marie Curie Career Integration Program. F. Z. and T. Z. acknowledge the financial support by the DFG Priority Program SPP1391. K. W. and Edwin Pun would like to thank the support from Research Grant Council of Hong Kong under Projects HKUST2/CRF/11G and AoE/P-02/12.

# References

1. Y. R. Shen, The principles of nonlinear optics, John Willey & Sons, New York, USA 1991.
2. R. W. Boyd, Nonlinear Optics, Academic Press, San Diego, USA 2008.
3. M. Kauranen and A. V. Zayats, Nature Photon. **6**, 737-748 (2012).
4. M. W. Klein, C. Enkrich, M. Wegener and S. Linden, Science **313**, 502-504 (2006).
5. W. Fan, S. Zhang, N. C. Panoiu, A. Abdenour, S. Krishna, R. M. Osgood, Jr., K. J. Mallo, and S. R. J. Brueck, Nano Lett. **6**, 1027-1030 (2006).
6. S. Kujala, B. K. Canfield, M. Kauranen, Y. Svirko, and J. Turunen, Phys. Rev. Lett. **98**, 167403 (2007).
7. R. Czaplicki, H. Husu, R. Siikanen, J. Makitalo, and M. Kauranren, Phys. Rev. Lett. **110**, 093902 (2013).
8. V. K. Valev, A. V. Silhanek, N. Verellen, W. Gillijns, P. Van Dorpe, O. A. Aktsipetrov, G. A. E. Vandenbosch, V. V. Moshchalkov, and T. Verbiest, Phys. Rev. Lett. **104**, 127401 (2010).
9. V. K. Valev, A. V. Silhanek, W. Gillijns, Y. Jeyaram, H. Paddubrouskaya, A. Volodin, C. G. Biris, N. C. Panoiu, B. De Clercq, M. Ameloot, O. A. Aktsipetrov, V. V. Moshchalkov, T. Verbiest, ACS Nano **5**, 91-96 (2011).
10. Y. Zhang, N. K. Grady, C. Ayala-Orozco, and N. J. Halas, Nano Lett. **11**, 5519-5523 (2011).
11. K. D. Ko, A. Kumar, K. H. Fung, R. Ambeka, G. L. Liu, N. X. Fang, and K. C. Toussaint, Nano Lett. **11**, 61-65 (2011).
12. W. Cai, A. P. Vasudev, and M. L. Brongersma, Science **333**, 1720-1723 (2011).
13. H. Aouani, M. N. Cia, M. Rahman, T. P. H. Sidiropoulos, M. Hong, R. F. Oulton, and S. A. Maier, Nano Lett. **12**, 4997-5002 (2012).
14. S. Linden, F. B. P. Niesler, J. Förstner, Y. Grynko, T. Meier, and M. Wegener, Phys. Rev. Lett. **109**, 015502 (2012).
15. M. Hentschel, T. Utikal, H. Giessen, and M. Lippitz, Nano Lett. 12, 3778–3782 (2012).
16. T. Utikal, T. Zentgraf, T. Paul, C. Rockstuhl, F. Lederer, M. Lippitz, and H. Giessen, Phys. Rev. Lett. **106**, 133901 (2011).
17. T. Hanke, G. Krauss, D. Trautlein, B. Wild, R. Bratschitsch, and A. Leitenstorfer, Phys. Rev. Lett. **103**, 257404 (2009).
18. H. Liu, G. X. Li, K. F. Li, S. M. Chen, S. N. Zhu, C. T. Chan, and K. W. Cheah, Phys. Rev. B **84**, 235437 (2011).
19. S. M. Chen, W. H. Wong, E. Y. B. Pun, K. W. Cheah, and G. X. Li, Adv. Opt. Mater. **1**, 22-526 (2013).
20. J. Renger, R. Quidant, N. van Hulst, and L. Novotny, Phys. Rev. Lett. **104**, 046803 (2010)
21. P. Genevet, J. P. Tetienne, E. Gatzogiannis, R. Blanchard, M. A. Kats, M. O. Scully, and F. Capasso, Nano Lett. **10**, 4880 (2010).

22. A. Rose, D. A. Powell, I. V. Shadrivov, D. R. Smith, and Y. S. Kivshar, Phys. Rev. B **88**, 195148 (2013).
23. O. E. Alon, V. Averbukh, and N. Moiseyev, Phys. Rev. Lett. **80**, 3743 (1998).
24. O. E. Alon, Phys. Rev. A **66**, 013414 (1992).
25. P. Zdanska and V. Averbukh, N. Moiseyev, J. Chem. Phys. **118**, 8726 (2003).
26. W. K. Burns and N. Bloembergen, Phys. Rev. B **4**, 3437 (1971).
27. S. Bhagavantam and P. Chandrasekhar, Proceedings of the Indian Academy of Sciences A **76**, 13 (1972).

FIGURES

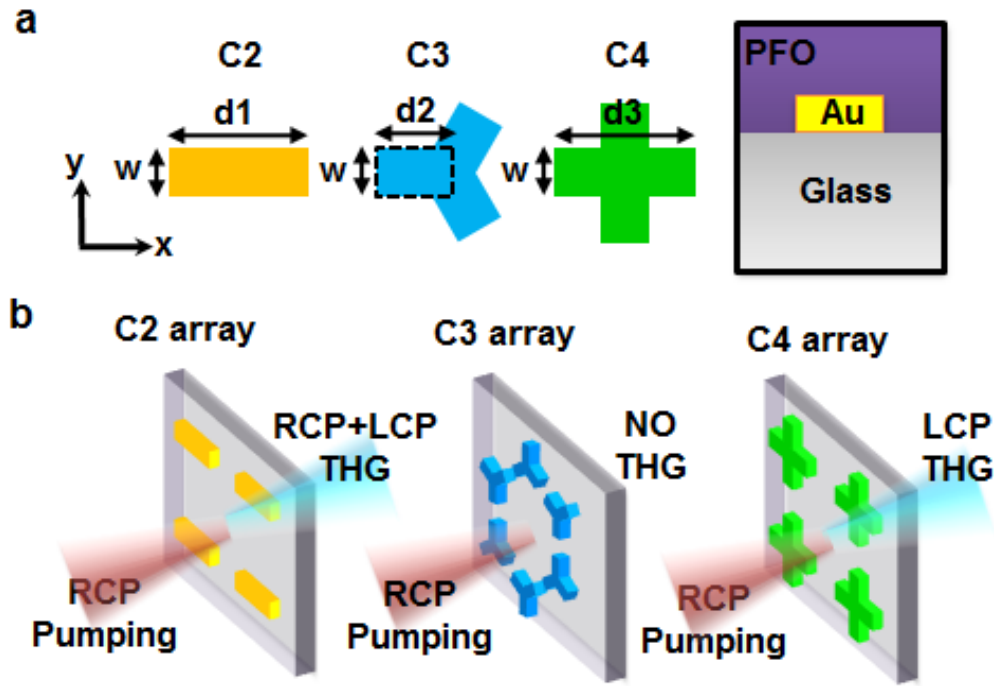

**Figure 1. Schematic of THG with nanostructures of various rotation symmetries.** (a) Geometry of plasmonic crystals with C2, C3 and C4 rotational symmetries. The unit cell consists of gold nanorods embedded in a thin PFO film. The width of the nanorods is w = 50 nm, and the lengths of nanorods are $d_1$=230 nm, $d_2$ = 110 nm and $d_3$ = 210 nm, respectively. The center to center distance between two adjacent unit elements is 400 nm with thickness of gold structures of t = 30 nm. (b) Illustration of the selection rule for right hand circularly polarized (RCP) light at the fundamental frequency for the gold-PFO hybrid crystals with two- (C2), three-(C3) and four-fold (C4) symmetry. The C2 symmetry sample generates both right- (RCP) and left- (LCP) circularly polarized third harmonic generation whereas the C3 symmetry sample does not produce any THG signal. Furthermore, the C4 symmetry sample generates only the opposite circularly polarized THG.

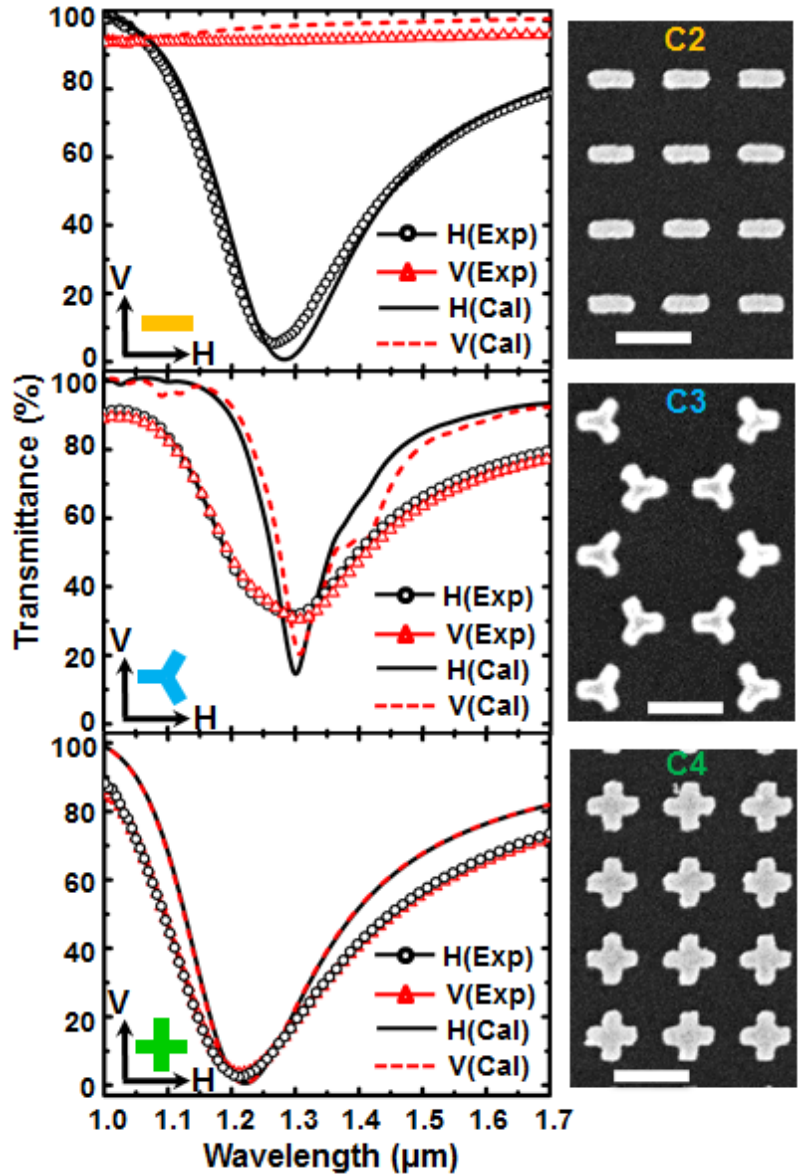

**Figure 2. Linear optical characterization of the metacrystals.** (Right) Scanning electron microscopy images of the gold nanorod samples with C2, C3 and C4 symmetry fabricated by electron beam lithography. The same bars represent 400 nm in all the three images. (Left) Measured (blue) and calculated (black) transmission spectra for the three kinds of gold-PFO plasmonic hybrid crystals showing the plasmonic resonance of the structures for illumination with horizontal (H-) and vertical (V-) polarized light (see inset for the orientation of the axes).

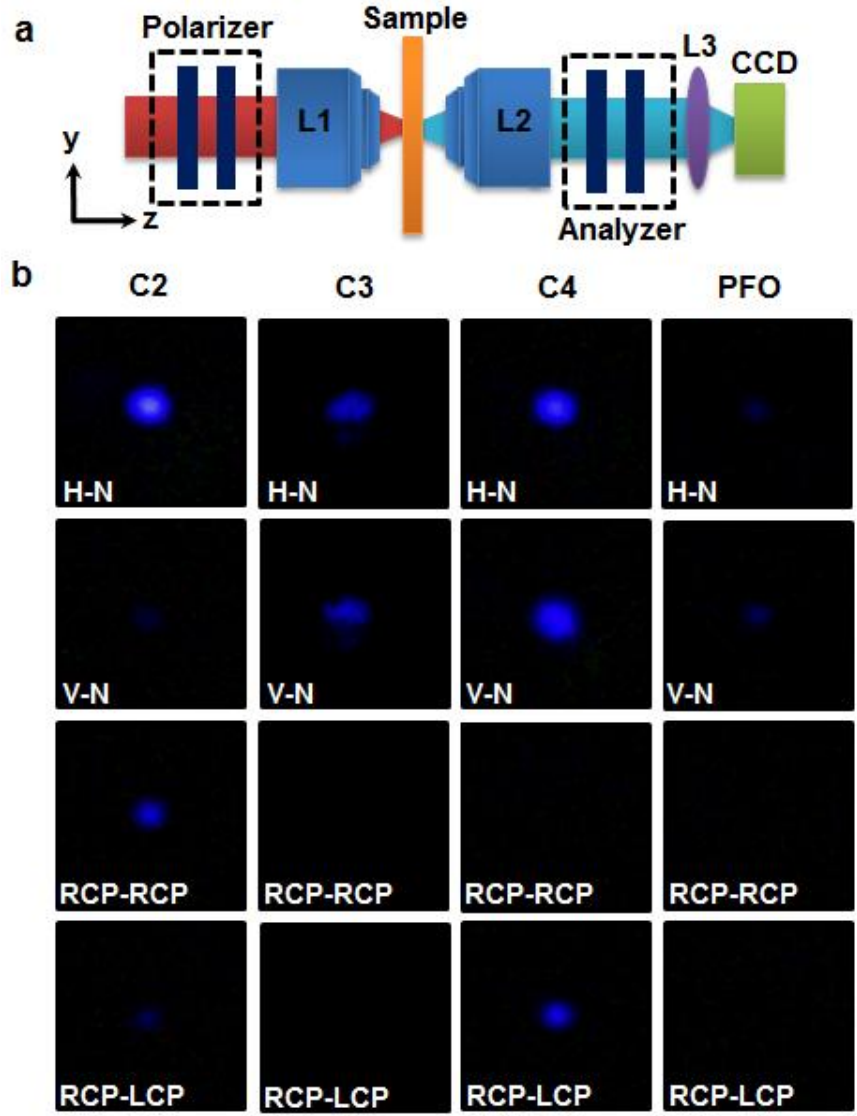

**Figure 3. Experimental results for the THG.** (a) Experimental setup for the THG measurement. The polarization of the pump light was controlled by the polarizer assembly which contains one linear polarizer and one quarter wave plate. The first objective lens (L1) was used to focus the light onto the plasmonic crystals. The second objective lens (L2) was used to collect the generated THG signals. Finally, the THG signal was imaged by a planar convex lens to a CCD camera. (b) Measured intensities of the THG signal from the samples with C2, C3, C4 symmetry as well as a bare PFO film for different polarization states of pump light. The first two rows show the results for pumping with linear polarization: horizontal (H) and vertical (V), receptively. 'N' means that no polarizer is used for the THG signal collection. The THG signals of the third and fourth rows were recorded using right circularly polarized pump light. The third row corresponds to the right circularly polarized component of the third harmonic signal whereas the fourth row shows the left circularly polarized component.

**Table 1.** Normalized intensity of the THG measured on C2, C3 and C4 symmetry metacrystals. 'N' means no polarizer ('-N') is used for THG signal collection, '-' means the signal is too weak to be observed.

|  | C2 | C3 | C4 |
|---|---|---|---|
| **H-N** | 1 | 1 | 1 |
| **V-N** | 0.02 | 0.70 | 0.91 |
| **R-R** | 0.18 | - | 0.01 |
| **R-L** | 0.02 | - | 0.30 |